\documentclass[12pt,a4paper,final]{iopart}

\usepackage{iopams} 
\usepackage{braket} 
\usepackage{graphicx}
\usepackage[breaklinks=true,colorlinks=true,linkcolor=blue,urlcolor=blue,citecolor=blue]{hyperref}
\usepackage[utf8]{inputenc}
\usepackage[T1]{fontenc}

\begin{document}

\title[Quantum repeater architecture with hierarchically optimized memory buffer times]{Quantum repeater architecture with hierarchically optimized memory buffer times}

\author[cor`]{Siddhartha Santra}
\address{US Army Research Laboratory, Adelphi, Maryland 20783, USA}
\ead{\mailto{sidsantra1@mgail.com}}

\author{Liang Jiang}
\address{Departments of Applied Physics and Physics, Yale University, New Haven, Connecticut 06520, USA}
\address{Yale Quantum Institute, Yale University, New Haven, Connecticut 06511, USA}
%\ead{author.two@mail.com}

\author{Vladimir S. Malinovsky}
\address{US Army Research Laboratory, Adelphi, Maryland 20783, USA}
%\eads{\mailto{author.three@mail.com}, \mailto{author.three@gmail.com}}

\begin{abstract}
We propose a quantum repeater protocol and architecture that mitigates decoherence of the entangled states by optimizing the quantum memory buffer time. The protocol maximizes the rate of distillable entanglement in the average accessed state at all nesting levels. The achievable rate is higher by orders of magnitude in comparison to a canonical protocol that does not optimize the buffer time. The advantage of the proposed design is observed for all nesting levels of the repeater for technologically feasible memory quality, entanglement generation and swapping success probabilities.
\end{abstract}

\vspace{2pc}
\noindent{\it Keywords}: Quantum repeaters, memory decoherence, optimized architecture.
%\maketitle

\section{Introduction}

Spatially distributed entanglement is a valuable resource for quantum communication, computing and sensing \cite{vanmeter}. Quantum repeaters (QR) follow a nested divide and conquer strategy to distribute entanglement across large distances \cite{briegel,dlcz}. At each nesting level, first, entangled states are generated probabilistically over smaller segments and stored in quantum memories at repeater stations. Second, a swapping operation on the memories doubles the physical range of the entangled state. As states make their way up the levels they spend some time, the memory buffer time, in the decohering quantum memories before being discarded or accessed for use by the next level. A larger buffer time at any nesting level increases the probability to obtain an entanglement length doubled state but decreases the entanglement quality of the average obtained state due to decoherence. These competing factors determine the entanglement generation rate (EGR) which is the product of the rate of obtaining entanglement length-doubled states and the entanglement of the average obtained state. An optimal buffer time maximizes the EGR. However, most protocols for repeater operation \cite{qrep-review-gisin} ignore the optimality of buffer time arising due to the interplay of entanglement generation probability and quantum memory decoherence \cite{imperfectmemory1}.\\
\indent In practice, it is crucial to include quantum memory decoherence for quantum repeaters that rely on two-way communication over long distances as shown in \cite{memerrors-hartmann}. The same reference suggests using decoherence free subspaces or local encoding and repeater operation in blind-mode to suppress memory errors. Other interesting ideas to address this challenge, e.g. addition of more physical resources such as multiplexed quantum memory to reduce memory waiting time \cite{multiplexed-memory}; or more complicated operations such as quantum error correction to actively suppress all errors \cite{qrep-encoding-jiang,qcomm-withoutmems,qcomm-FT} are promising in the long term but still very challenging with current experimental capability. Besides asking for more physical resources or complicated operations, it is also important to optimize the parameters of QR protocols. For example, dynamic programming has been introduced to explore the huge parameter space of QR protocols, which can successfully identify efficient protocols with significantly boosted performance in the absence of memory decoherence \cite{Jiang17291}. So far, there is no efficient method that can include quantum memory decoherence and systematically optimize the design parameters of QR protocols.\\
\indent Here, we propose an optimized buffer time protocol (OBP) and architecture that maximizes the entanglement generation rate using hierarchically optimized buffer times for all nesting levels. The optimal buffer time depends on the parameters of quantum memory quality, $\beta$, the entanglement generation probability, $p$, and the swapping success probability, $p_S$. The minimal parametrization chosen in terms of $(p,\beta,p_S)$ subsumes implementation-specific details such as source-station geometry, coupling and conversion efficiences or the use of multiplexed memories etc. (Entanglement generation probability $p$, for example, can include source-fiber coupling, wavelength conversion and memory read-in efficiency. Memory read-out may be included in $p$ or swapping success probability $p_S$.) We compare the OBP to a canonical repeater protocol (CP) that does not optimize the buffer time and show that the OBP improves the entanglement generation rate by several orders of magnitude in the technologically relevant parameter region. Moreover, we show that the relative improvement due to the OBP increases with the nesting level for technologically feasible swapping success probability. The protocol works for finite-lifetime quantum memories used to store entangled states in quantum repeaters which utilize two-way classical communication between its nodes to verify entanglement generation before entanglement swapping is performed.

The layout of the paper is as follows. Section \ref{sec:repprotocol} first describes the central idea of the optimized memory buffer time protocol in subsection \ref{subsec:optfirnest} followed by the definition of optimal memory buffer time in subsection \ref{subsec:optaccesstime} and comparison with a canonical protocol in subsection \ref{subsec:egrcomp}. Section \ref{sec:arch} presents a quantum repeater architecture compatible with hierarchical optimization of buffer times in subsection \ref{subsec:hierarch}. Subsection \ref{subsec:hieropt} then describes an algorithm that can be used for the hierarchical optimization. Further, subsection \ref{subsec:compallnest} shows the comparison of the entanglement generation rates for the OBP compared to the CP for all nesting levels. Section \ref{sec:conc} concludes the paper with a discussion of the typical advantage one may expect using OBP when used with state of art parameters.

%%%%%%%%%%%%%%%%%%%%%%%%%%%%%%%%%%%%%%%%%%%%%%%%%%%%
\section{Quantum repeater protocol with optimized memory buffer time}
\label{sec:repprotocol}
In this section we describe the optimized memory buffer time protocol by focusing on the first nesting level, in subsection \ref{subsec:optfirnest}, of a potentially multi-level quantum repeater network. While quantum memories may  also suffer from decoherence due to depolarization and  loss, we consider dephasing as the only mode of memory decoherence.  
This highlights the central physical idea while keeping the discussion mathematically simple. The optimal buffer time is described in subsection \ref{subsec:optaccesstime}. A comparison of entanglement generation rates of the optimized and canonical protocols is presented in subsection \ref{subsec:egrcomp}. 

We term a repeater protocol that does not optimize its quantum memory buffer time as a canonical protocol, for example, those in \cite{imperfectmemory1} and \cite{memerrors-hartmann}. As with the optimized protocol, in canonical protocols entanglement generation and swapping occur probabilistically. For comparison with the optimized protocol, the distinctive feature of canonical protocols is that the quantum memories can wait for arbitrarily long times for successful entanglement generation. The rate of entanglement generation in such protocols is inversely proportional to the expected number of entanglement generation attempts needed for success, as shown in \ref{app:avstates}. Subsequent to entanglement generation, for both the canonical and optimized protocols, purification of the generated entangled pairs may be performed if multiple quantum memories at a given nesting level are available at the repeater nodes. In this paper, we compare the protocols without considering purification of entangled states on a finite number of quantum memories. Thus, we compare the two protocols on a single-copy basis and use the distillable entanglement of the average state in the respective protocols as a measure of the entanglement quality.

%%%%%%%%%%%%%%%%%%%%%%%%%%%%%%%%%%%%%%%%%%
\subsection{Optimized memory buffer time protocol at the first nesting level}
\label{subsec:optfirnest}
\begin{figure}[h]
\centering
\includegraphics[height=6cm,width=.6\columnwidth]{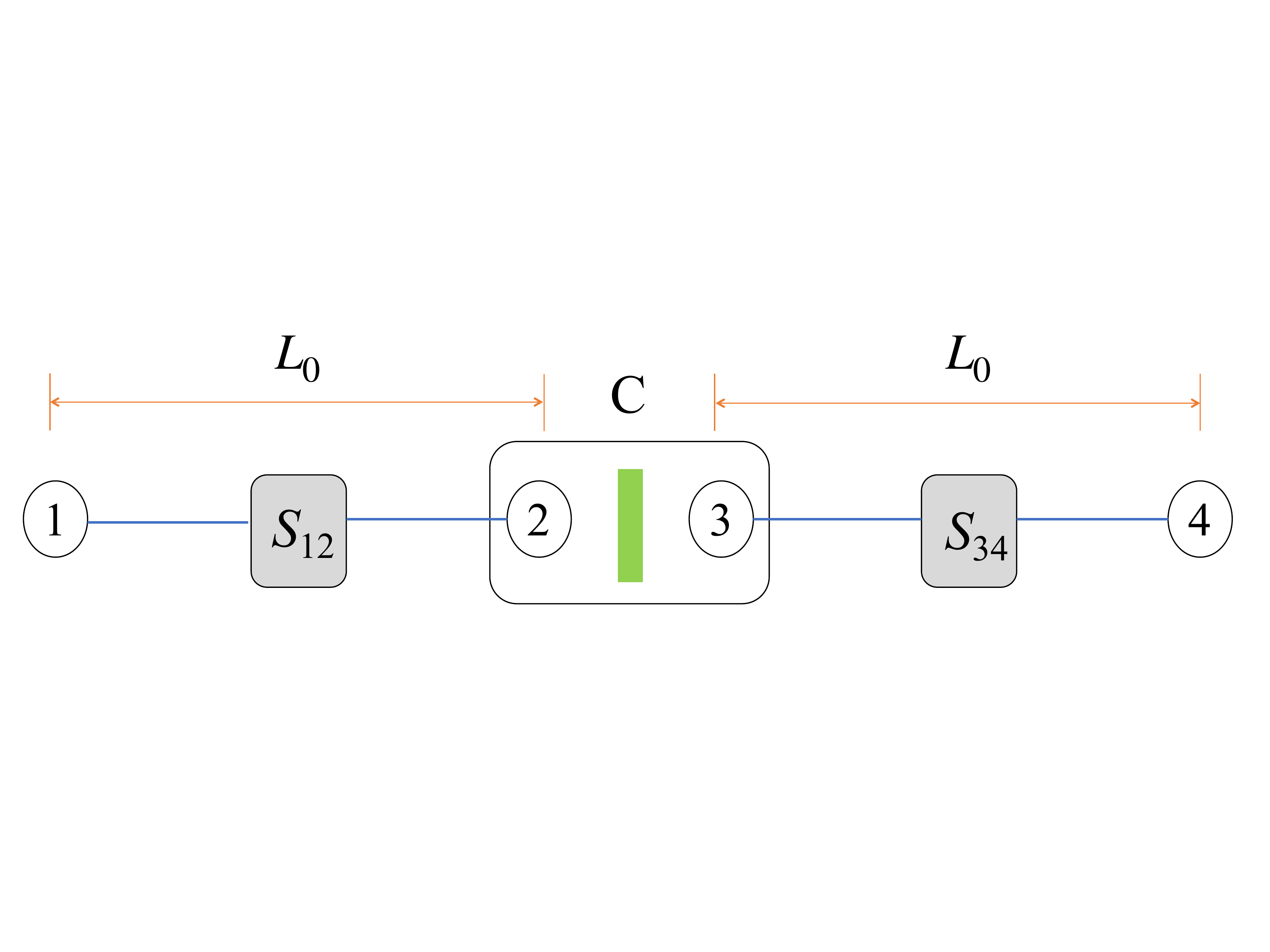}
%\vspace{-1cm}
\caption{(Color online) A quantum repeater with two segments at nesting level 1. Quantum memory pairs $1,2$ and $3,4$ store entangled states produced by sources $S_{12}$ and $S_{34}$ for entanglement swapping, green rectangle, at repeater station $C$.}
\label{fig:qlink}
\end{figure}
\noindent Operationally, the optimized memory buffer time protocol can be understood by considering the entanglement swapping of states across two elementary segments at the first nesting level in a QR, Figure~\ref{fig:qlink}. Sources $S_{12}$ and $S_{34}$ supply entangled states with probability, $p$, to the decohering quantum memory pairs $(1,2)$ and $(3,4)$. The memory lifetime is denoted by $\tau_M$. Entanglement swapping at the repeater station, C, is performed via informed Bell-state measurements. C checks the two pairs of memories verifying whether they are charged which requires waiting for one unit of one-way classical communication time, $\tau_C=L_0/c$, with $c$ the speed of light in the fiber and $L_0$ the length of the segment. If both pairs are charged, C performs a swapping operation on the memories $2$ and $3$ with success probability, $p_S$, producing an entangled state across the remote memories $1$ and $4$. Information about the success or failure of the swapping operation is then communicated to the remote memories taking an additional time $\tau_C$. In case C finds the memory pairs uncharged (one or both) it classically communicates the need to continue entanglement generation attempt in the segment(s) to the remote memories which also requires $\tau_C$ amount of time. The OBP limits the number of such entanglement generation attempts to a number $n_{\textrm{opt}}(p,\beta,p_S)$, determined by the operating parameters, after which the state from the remote memories is accessed. Subsequently all four memories are refreshed and the entanglement generation process starts over. The CP on the other hand places no limit on the number of entanglement generation attempts which continue till a state is obtained in both segments.\\

By limiting the buffer time, the OBP provides an average remote entangled state with a high measure of entanglement since the time for memory decoherence is limited. The entanglement generation in the two segments can succeed at step numbers, $1\leq k_1,k_2\leq n$, where $n$ is the maximum number of attempts and the probability distribution of successful entanglement generation is $\mathbb{P}(k_1,k_2)=(1-p)^{k_1+k_2-2}p^2$. Without loss of generality, we assume that the memory pairs in the two segments of figure~\ref{fig:qlink} are supplied with the state, $\rho^-=\ket{\psi^-}\bra{\psi^-}$, where, $\ket{\psi^\pm}=(\ket{01}\pm\ket{10})/\sqrt{2}$.  Storing the $\rho^-$ state in a pair of quantum memories with lifetime $\tau_M$ for time $t$ results in the state, $\rho(t)=\rho^-(1+e^{-2t/\tau_M})/2+\rho^+(1-e^{-2t/\tau_M})/2$, where $\rho^+=\ket{\psi^+}\bra{\psi^+}$. The remote state obtained after a successful swap and communication to the remote memories $1$ and $4$ is
\begin{equation}
\rho^S(k_1,k_2)=\frac{1}{2}(1+\beta^{|\Delta k|+2})\rho^-+\frac{1}{2}(1-\beta^{|\Delta k|+2})\rho^+,
\label{swapstategreedy}
\end{equation}
where $\beta=e^{-2\tau_C/\tau_M}$ is the memory quality parameter that quantifies the decoherence in a pair of quantum memories during one round of one-way classical communication,  and $\Delta k =(k_2-k_1)$. $\rho^S(k_1,k_2)$ approaches the totally mixed state exponentially fast with $\Delta k$. Thus, the states for large $\Delta k$ contribute little to the entanglement of the average state. The state in Eq.~(\ref{swapstategreedy}) further decoheres in the remote memories for a time, $t=2\tau_C(n-\textrm{max}(k_1,k_2))$, before being accessed, and leads to the state $\rho^S_n(k_1,k_2)=\rho^-(1+\beta^{|\Delta k|+2+n-\textrm{max}(k_1,k_2)})/2+\rho^+(1-\beta^{|\Delta k|+2+n-\textrm{max}(k_1,k_2)})/2$. The average remote entangled state is the probabilistically weighted sum of such states, $\rho^O= \mathcal{N}^{-1} \sum_{k_1=1,k_2=1}^{n,n}  \mathbb{P}(k_1,k_2)\rho^S_n(k_1,k_2)$, where $\mathcal{N}=(1-(1-p)^n)^2$ is the total probability of obtaining a remote entangled state across the two segments in $n$ attempts. The average remote entangled state can be expressed in a compact manner as (see \ref{app:avstates}),
\begin{equation}
\rho^O=\frac{1}{2}(1+\gamma^O(p,\beta,n))\rho^-+\frac{1}{2}(1-\gamma^O(p,\beta,n))\rho^+,
\label{avstateopt}
\end{equation}
which has a fidelity of $F^O(p,\beta,n)=\tr\{\rho^-\rho^O\}=\frac{1}{2}(1+\gamma^O(p,\beta,n))$ and is obtained once every $n 2\tau_C$ period of time. The function $\gamma^O(p,\beta,n)\in[0,1]$ can be physically interpreted as the degradation in fidelity of the average state due to memory decoherence during the buffer time. For perfect quantum memories, $\beta=1$, and $\gamma^O(p,\beta,n)=1$.

%%%%%%%%%%%%%%%%%%%%%%%%%%%%%%%%%
\subsection{Optimal memory buffer time}
\label{subsec:optaccesstime}

As an entanglement measure for the mixed state, $\rho^O$, we use the upper bound on its distillable entanglement, $E[F^O]=H[\frac{1}{2}+(F^O(1-F^O))^{.5}]$ for $1\geq F^O>0.5$ and $E[F^O]=0$ for $.5\geq F^O\geq 0$, where $H[x]=-x\log(x)-(1-x)\log(1-x)$ is the binary entropy function. $E[F^O]$ expresses the number of pure Bell states that the best distillation protocol can achieve in the limit of asymptotic number of copies. For $\rho^O$ the bound can be achieved using the hashing protocol \cite{entdistill}. The entanglement generation rate is thus given by the rate of distillable entanglement (DE),
\begin{equation}
R^O_{DE}(p,\beta,n)=\frac{p_S(1-(1-p)^n)^2}{n(2\tau_C)}E[F^O(p,\beta,n)].
\label{rateopt}
\end{equation}
The optimal buffer time, $n_{\textrm{opt}}$, maximizes this rate for given $p$  and $\beta$ values, i.e.,
\begin{equation}
n_{\textrm{opt}}(p,\beta)=\textrm{ArgMax}_{n}[R^O_{DE}(p,\beta,n)].
\label{eq:noptimal}
\end{equation}
with the obvious condition that $n_{\textrm{opt}}(p,\beta)\geq 1$. Note that the optimal buffer time found using (\ref{eq:noptimal}) is obtained in units of the two-way classical communication time $2\tau_C$. The behavior of the optimal buffer time, $n_{\textrm{opt}}(p,\beta)$, in different regions of the $(p,\beta)$ parameter space is described in \ref{app:nopt}. While the optimal buffer time at the first nesting level depends only on the entanglement generation probability $p$ and the memory quality parameter $\beta$, for higher nesting levels it depends also on the swapping success probability of the previous level $p_S$. In this paper, for simplicity, we assume that the parameters $p,\beta,p_S$ remain constant for all nesting levels. However, our analysis outlined in section \ref{sec:arch} can be used to address various distributions of the parameters. In case of multiplexed quantum memories \cite{multiplexed-memory}, expression (\ref{eq:noptimal}) can be used to determine the optimal buffer time by using the effective entanglement generation probability between the nodes.
Also, note that the asymptotic value of the fidelity, $F^O(p,\beta,n)=\frac{1}{2}(1+\gamma^O(p,\beta,n))$, is at least $0.5$ when dephasing is the only mode of decoherence. Indeed, dephasing is the dominant mode of decoherence for repeater-relevant timescales in quantum memories based on nuclear spins in diamond NV centers and the hyperfine electron levels in ion traps \cite{Simon2010}. However, when loss and depolarization are also considered, the asymptotic fidelity can fall below the distillable entanglement threshold of $F^O>0.5$. The optimized memory buffer time protocol works in this general case as well but now the maximum memory buffer time, $n_{\textrm{max}}$, is limited by the threshold condition, $F^O(p,\beta,n_{\textrm{max}})>0.5$. 
%%%%%%%%%%%%%%%%%%%%%%%%%%%%%%%%%%
\subsection{Entanglement generation rate comparison of the optimized and canonical protocols}
\label{subsec:egrcomp}

\begin{figure}
\centering
\includegraphics[height=7cm,width=.6\columnwidth]{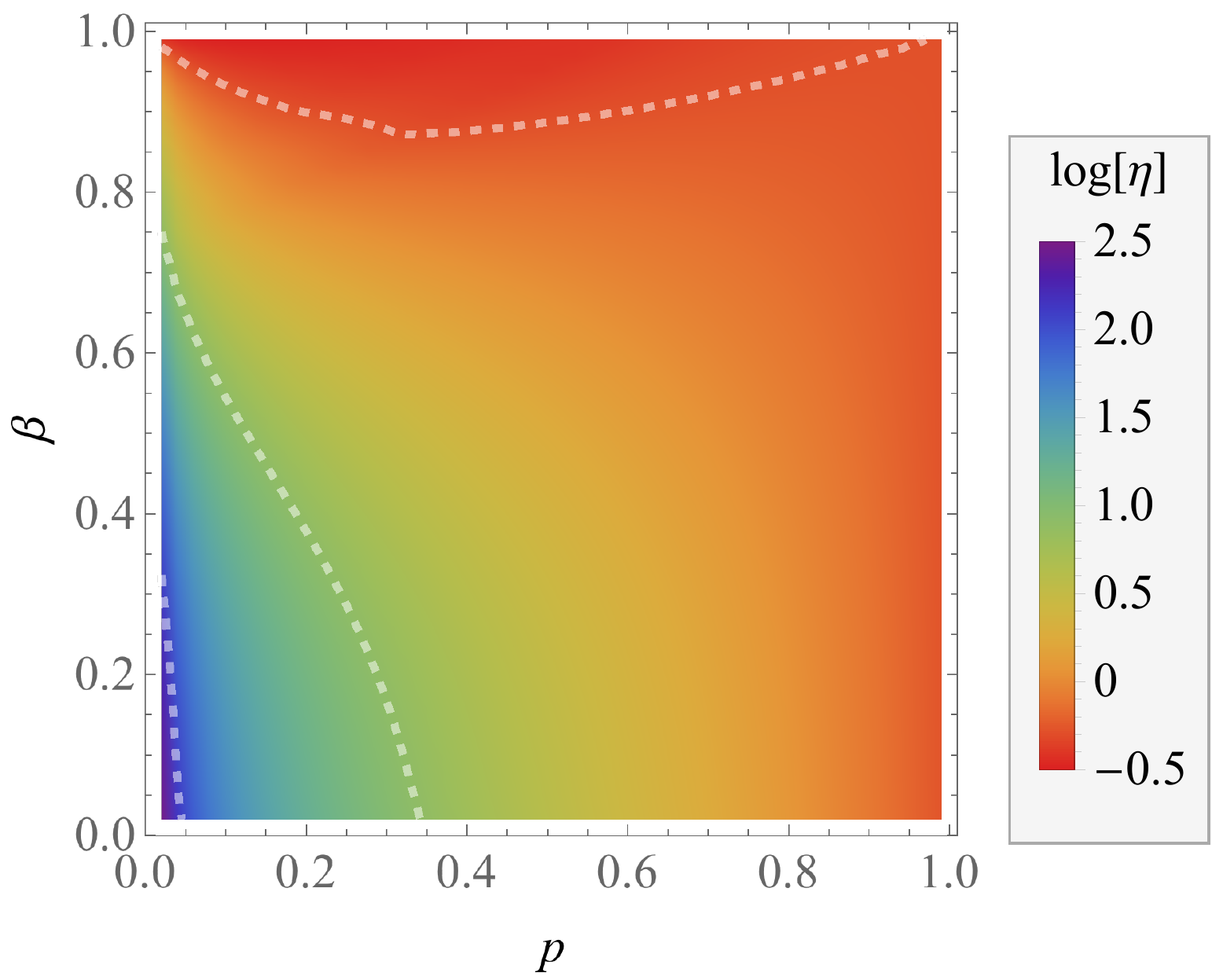}
\caption{(Color online) Logarithm (base 10) of the ratio of entanglement generation rate in the optimized memory buffer time protocol, $R^O_{DE}$, to that in the canonical protocol, $R^C_{DE}$, at the first nesting level. Dotted white lines are contours with $\log[\eta]=2,1,0$ from left to right.}
\label{fig:rateratio-p-beta1}
\end{figure}

To compare the entanglement generation rates of the optimized and canonical protocols we next obtain the average remote entangled state of the canonical protocol. The average remote entangled state in the canonical protocol has a low measure of entanglement since it is an average over states that have decohered in the memories for arbitrarily long times. The average state, $\rho^C= \sum_{k_1=1,k_2=1}^{\infty,\infty}  \mathbb{P}(k_1,k_2)\rho^S(k_1,k_2)$, again takes a compact form (see \ref{app:avstates})
\begin{equation}
\rho^C=\frac{1}{2}(1+\gamma^C(p,\beta))\rho^-+\frac{1}{2}(1-\gamma^C(p,\beta))\rho^+.
\label{avstatecan}
\end{equation}
Such states of fidelity $F^C(p,\beta):=\tr\{\rho^-\rho^C\}=\frac{1}{2}(1+\gamma^C(p,\beta))$ are obtained at the rate of the inverse of the waiting time $\braket{k}=(3-2p)/p(2-p)$ \cite{rate-vanloock}. The entanglement generation rate in the canonical protocol is 
\begin{equation}
R^C_{DE}(p,\beta)=\frac{p_S}{\braket{k}(2\tau_C)}E[F^C(p,\beta)].
\label{ratecan}
\end{equation}

The optimized buffer time protocol provides manifold increase of entanglement generation rates in most of the $(p,\beta)$-parameter space, even at the first nesting level, as shown in figure~\ref{fig:rateratio-p-beta1}. In particular, for the low $p,\beta$-region the ratio, $\eta(p,\beta,n_{\textrm{opt}}) = R^O_{DE}(p,\beta,n_{\textrm{opt}})/R^C_{DE}(p,\beta)\sim1/p$ (see \ref{app:rateratio}). Only for $\beta\sim 1$, i.e., for near-perfect quantum memories, does the canonical protocol provide better rates. The optimal buffer time, $n_{\textrm{opt}}$, depends on the operating point in parameter space. For short-lived quantum memories, $\beta<<1$, it is numerically found that $n_{\textrm{opt}}=1$. For long-lived quantum memories, $\beta\to 1$, and low entanglement generation probability, $p\to 0$, the optimal buffer time scales as, $n_{\textrm{opt}}\sim p^{-1}\log(1/\beta)^{-1}=(1/p)(\tau_M/2\tau_C)$.

%%%%%%%%%%%%%%%%%%%%%%%%%%%%%%%%%%%%%%%%%%%%%%%%%%%%%%%
\section{Hierarchical buffer time optimization-compliant repeater architecture}
\label{sec:arch}

We now present a repeater architecture which can operate all its nesting levels based on the optimized memory buffer time protocol in subsection \ref{subsec:hierarch}. This is followed by a description of the algorithm to hierarchically optimize the buffer time in subsection \ref{subsec:hieropt}. The section ends by presenting a comparison of the entanglement generation rates of the optimized and canonical protocols for all nesting levels in subsection \ref{subsec:compallnest}. 

%%%%%%%%%%%%%%%%%%%%%%%%%%%%%%%%%%%%%
\subsection{Optimization-compliant architecture for all nesting levels}
\label{subsec:hierarch}
\begin{figure}[h]
\centering
\vspace{-.5cm}
\includegraphics[height=9cm,width=.9\columnwidth]{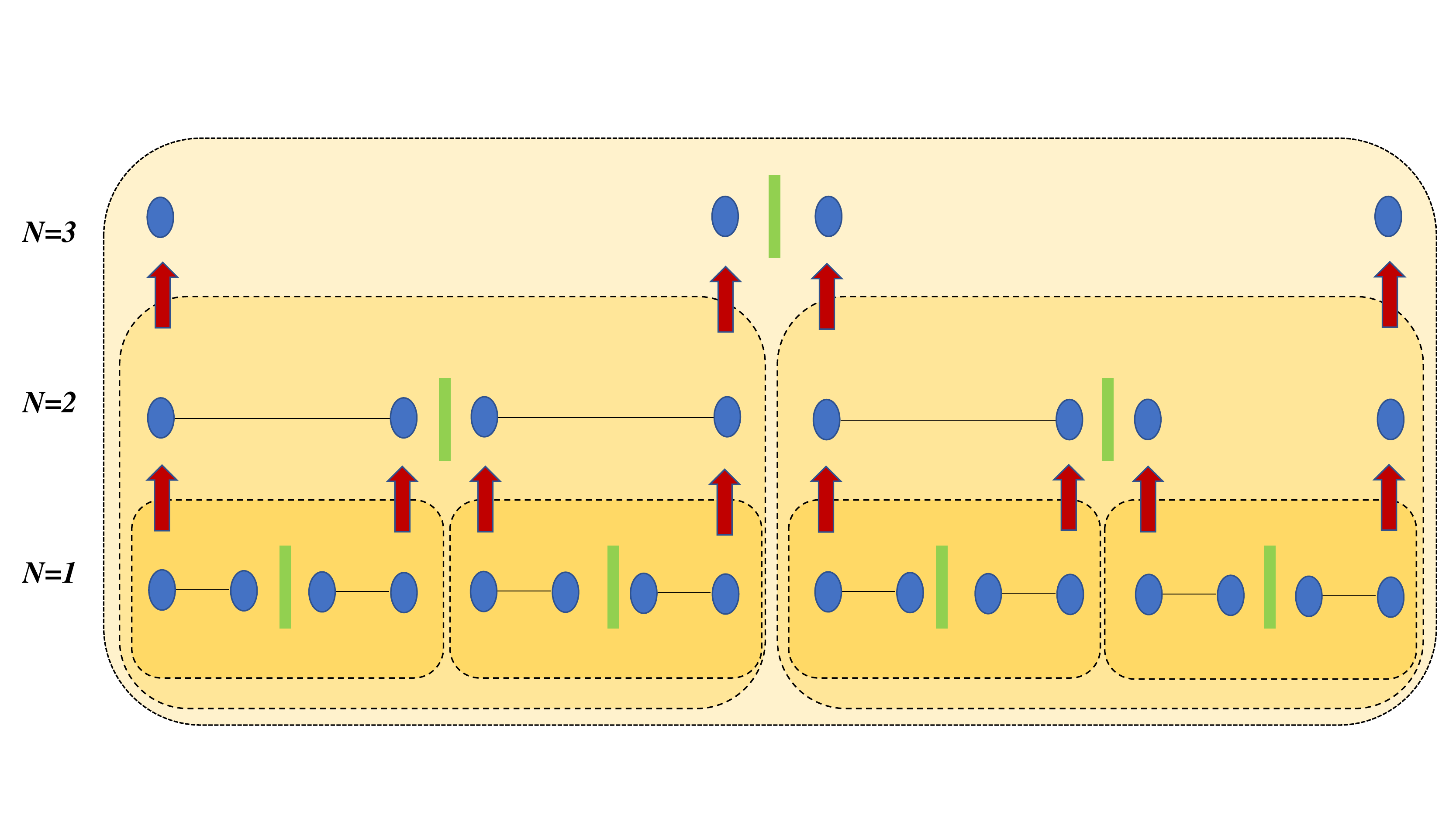}
\vspace{-2cm}
\caption{(Color online) Quantum repeater architecture based on the optimized buffer time protocol. Shown are three nesting levels ($i=1,2,3$) and 9 repeater nodes ($A-I$). A new set of quantum memories (blue ovals) are required at each nesting level. Green rectangles represent entanglement swapping operations. Entanglement length-doubled states obtained at any level are transferred to the quantum memories of the next level using coherent operations (red arrows).}
\label{fig:nestedlevels}
%\vspace{-.5cm}
\end{figure}
\noindent A quantum repeater architecture capable of supporting hierarchical optimization of buffer times requires a new set of quantum memories at each nesting level as shown in figure~\ref{fig:nestedlevels}. The average remote entangled state output by nesting level $i$ is transferred to the new quantum memories at level $(i+1)$ for $i=1,2,...,(N_m-1)$, where $N_m$ is the maximum nesting level. This transfer can be achieved by using a two-qubit quantum SWAP gate \cite{swap-experiment}. In figure~\ref{fig:nestedlevels} the quantum memories are labeled by the nesting level as superscript and the node label as subscript, for example, $m^{(1)}_{E_1}$ denotes quantum memory number 1 at the first nesting level of node $E$. The memories at any level follow the OBP with an optimal buffer time that is determined by the effective probability with which it receives entangled states and memory quality parameter relative to its classical communication time. All levels follow the informed Bell-state measurement procedure followed by communication to the remote memories just as in the first nesting level described earlier. The state output by a nesting level is therefore the probabilistically weighted sum of the average state received from the previous level. Periodic SWAP operations between two quantum memories at a node, for example, between $m^{(1)}_{E_1} \to m^{(2)}_{E_1} \to m^{(3)}_{E_1}$ in figure~\ref{fig:nestedlevels}, are used to feed forward the average state to the quantum memories of the higher nesting level.

\begin{figure}[h]
\centering
\includegraphics[height=6cm,width=.8\columnwidth]{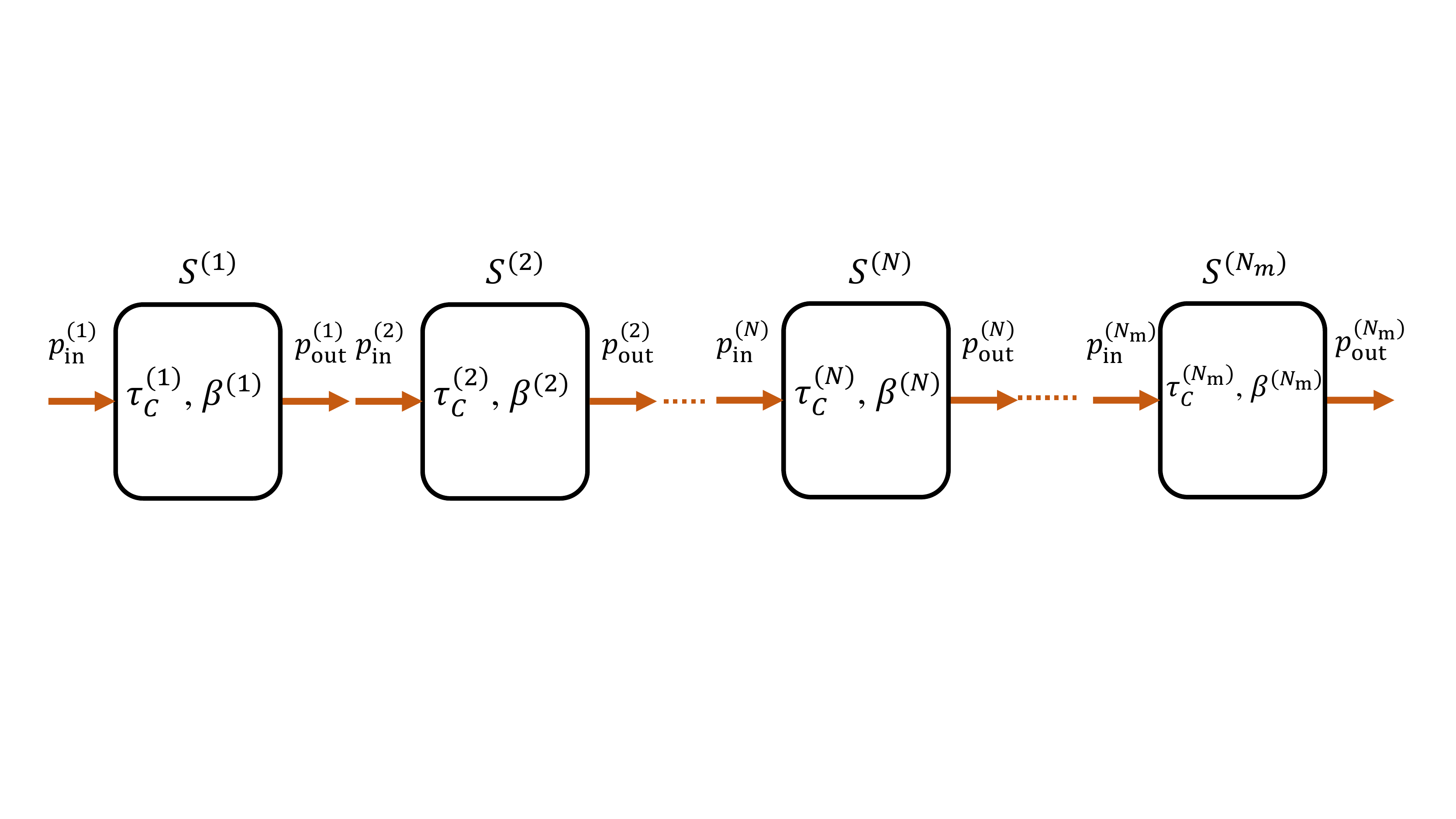}
\vspace{-2cm}
\caption{(Color online) Organization of nesting levels in a quantum repeater architecture as a sequence of self-similar systems. Each system $S^{(i)}$ is characterized by the one-way classical communication time $\tau_C^{(i)}$ and its memory quality parameter $\beta^{(i)}$. $S^{(i)}$ receives a state $\rho^{(i)}_{\textrm{in}}$ with probability $p^{(i)}_{\textrm{in}}$ every $n^{(i)}_{\textrm{in}}\tau^{(i)}_C$ amount of time from the preceding system $S^{(i-1)}$. $S^{(i)}$ outputs a state $\rho^{(i)}_{\textrm{out}}$ with probability $p^{(i)}_{\textrm{out}}$ every $n^{(i)}_{\textrm{out}}$ of its input cycles.}
\label{fig:in-out-system}
\end{figure}

The nesting levels in such an architecture can be modeled as a sequence of self-similar input-output systems, $\{S^{(i)}\}_i,i\in\{1,2,...,N_m\}$, as shown in figure~\ref{fig:in-out-system}. Each system is characterized by its classical communication time $\tau_C^{(i)}=2^{i-1}\tau_C$ and memory quality parameter $\beta^{(i)}=\beta^{2^{i-1}}$. Further, for each system $S^{(i)}$ the input-cycle time $n^{(i)}_{\textrm{in}}$ specifies the number of two-way classical communication cycles over which it receives one average state from the previous system with probability $p^{(i)}_{\textrm{in}}$. This takes $n^{(i)}_{\textrm{in}}2\tau^{(i)}_C$ amount of time. The output-cycle time $n^{(i)}_{\textrm{out}}$ is the buffer time for system $S^{(i)}$ in terms of the number of input-cycles for system $S^{(i)}$, i.e., one average state is output by the system in $n^{(i)}_{\textrm{out}}n^{(i)}_{\textrm{in}}2\tau^{(i)}_C$ amount of time with probability $p^{(i)}_{\textrm{out}}$. 
Any system $S^{(i)}$ is able to receive or output an average state only at the end of a time period that is a multiple of its two-way classical communication cycle time. Two adjacent systems $S^{(i)}$ and $S^{(i+1)}$ are synchronized if the physical times at which the $i$'th system outputs its average state corresponds to the physical times at which the $(i+1)$'th system can receive the state. Therefore, successive input-cycle times and output-cycle times have to obey the condition for synchronization of the systems, $n^{(i)}_{\textrm{out}}n^{(i)}_{\textrm{in}}2\tau_C^{(i)}=n^{(i+1)}_{\textrm{in}}2\tau^{(i+1)}_C$, which implies
\begin{equation}
n^{(i+1)}_{\textrm{in}}=n^{(i)}_{\textrm{out}}n^{(i)}_{\textrm{in}}/2,
\label{eq:phasematchcond}
\end{equation}
for $i=1,...,N_m-1$, so that  $n^{(i)}_{\textrm{in}, \textrm{out} }$ are positive integers, and $n^{(1)}_{\textrm{in}}=1$. The output probability of system $S^{(i)}$ is related to its input probability as $ p^{(i)}_{\textrm{out}}=p_s[1-(1-p^{(i)}_{\textrm{in}})^{n^{(i)}_\textrm{out}}]^2$ with $p^{(1)}_{\textrm{in}}=p$. While the input probability of system $S^{(i)}$ is related to the output probability of system $S^{(i-1)}$ as $p^{(i)}_{\textrm{in}}=p_Tp^{(i-1)}_{\textrm{out}}$, where $p_T$ is the probability to successfully transfer states from the memories of one nesting level to the next, and $ p^{(0)}_\textrm{out}=p$ to maintain consistency.

The average remote entangled state $\rho^{O,(i)}$ obtained in the OBP based architecture at any nesting level $i$ depends on the parameter values for the preceding levels, i.e., on $\{\tau^{(j)}_C,\beta^{(j)}_C,n^{(j)}_\textrm{in}\}$ for $1\leq j\leq i$. In addition, it depends on the initial entanglement generation probability, $p$, the swapping success probability, $p_S$, and the transfer success probability, $p_T$ . This state, 
\begin{equation}
\rho^{O,(i)}=\frac{1+\prod_{j=1}^i[\gamma^{O,(j)}]^{2^{i-j}}}{2}\rho^- +\frac{1-\prod_{j=1}^i[\gamma^{O,(j)}]^{2^{i-j}}}{2}\rho^+,
\label{eq:rhonestleveli}
\end{equation}
is obtained once every $n^{(i)}_{\textrm{in}}n^{(i)}_{\textrm{out}}2\tau^{(i)}_C$ amount of time. $\gamma^{O,(j)}$ as a function of the relevant parameters can be found in the supplementary material. Physically, $\gamma^{O,(j)}\leq 1$ can be understood as the degradation of the fidelity during the buffer time at nesting level $j$. $\rho^{O,(i)}$ has a fidelity $F^{O,(i)}=\tr\{\rho^-\rho^{O,(i)}\}=(1+\prod_{j=1}^i[\gamma^{O,(j)}]^{2^{i-j}})/2$. This fidelity  approaches the distillation threshold of $1/2$ as $(1/2)\prod_{j=1}^i[\gamma^{O,(j)}]^{2^{i-j}}$, which accounts for the degradation in all nesting levels prior to $i$. Therefore, the EGR of system $S^{(i)}$ is 
\begin{equation}
R^{O,(i)}_{DE}=\frac{p^{(i)}_{\textrm{out}}}{n^{(i)}_{\textrm{in}}n^{(i)}_{\textrm{out}}2\tau^{(i)}_C}E\left[\frac{1+\prod_{j=1}^{i}[\gamma^{O,(j)}]^{2^{i-j}}}{2}\right].
\label{DEratei}
\end{equation}

%%%%%%%%%%%%%%%%%%%%%%%%%%%%%%%%%%%%%%%%%%%
\subsection{Hierarchical optimization of memory buffer time at every nesting level}
\label{subsec:hieropt}

In a repeater with $N_m$ levels the final EGR can be maximized by hierarchically optimizing buffer times of all nesting levels. We search over the set of positive integers $\{n^{(i)}_{\textrm{in}},n^{(i)}_{\textrm{out}}\}$, subject to synchronization condition constraints, Eq.~(\ref{eq:phasematchcond}). The optimization proceeds by maximizing the EGR sequentially starting from nesting level 1. If at any nesting level, $i$, the optimal value for $n^{(i)}_{\textrm{out,opt}} = \textrm{ArgMax}_{n^{(i)}_{\textrm{out}}}[R^{O,(i)}_{DE}]$ does not satisfy the synchronization condition one of the neighboring values, $\tilde{n}^{(i)}_{\textrm{out,opt}}\in\{n^{(i)}_{\textrm{out,opt}}-1,n^{(i)}_{\textrm{out,opt}}+1\}$ whichever provides a higher $R^{O,(i)}_{DE}$, is chosen and  used for calculating $n^{(i+1)}_{\textrm{in}}$ and $p^{(i+1)}_{\textrm{in}}$. This procedure is followed for all nesting levels upto $i=N_m$ and produces an approximately-optimal synchronized sequence of $n^{(i)}_{\textrm{in}}$ and $n^{(i)}_{\textrm{out}}$. A repeater operating its nesting levels based on the sequence of approximately-optimal buffer times still gives manifold increase of EGR as shown by the logarithm of the ratio of the rates at nesting level $i$, $\log[\eta^{(i)}(p,\beta,n)]=\log[R^{O,(i)}_{DE}(p,\beta,n)/R^{C,(i)}_{DE}(p,\beta)]$, in figure~\ref{fig:crossover}. 

\subsection{Comparison of optimized vs canonical protocol at any nesting level}
\label{subsec:compallnest}

In the canonical protocol all nesting levels operate on the same set of quantum memories and successive levels do not use new set of memories. An average remote entangled state at any level is obtained after the waiting time for that level. The next level receives the average state from the previous level as soon as it is obtained, i.e., no synchronization conditions are used. The average remote entangled state in the canonical protocol obtained at any nesting level $i$ is given by an expression identical to Eq.~(\ref{eq:rhonestleveli}) with $\gamma^{C,(j)}$ replacing $\gamma^{O,(j)}$. Physically, $\gamma^{C,(j)}$ can be understood as the degradation in the fidelity due to the waiting time at nesting level $j$.  The entanglement generation rate in the canonical protocol is given by (see \ref{app:avstates})
\begin{equation}
R^{C,(i)}_{DE}=\frac{p_S}{(\prod_{j=1}^i\braket{k}_j)2\tau_C}E\left[\frac{1+\prod_{j=1}^{i}[\gamma^{C,(j)}]^{2^{i-j}}}{2}\right] ,
\label{canDEratei}
\end{equation}
where $\braket{k}_1=(3-2p)/p(2-p)$ is the waiting time at the first nesting level due to the initial entanglement generation probability, and $\braket{k}_j=(3-2p_S)/p_S(2-p_S)\forall j\geq 2$ is the waiting time due to the swapping success probability at the second nesting level and higher. 
\begin{figure}
\centering
\includegraphics[height=6cm,width=.6\columnwidth]{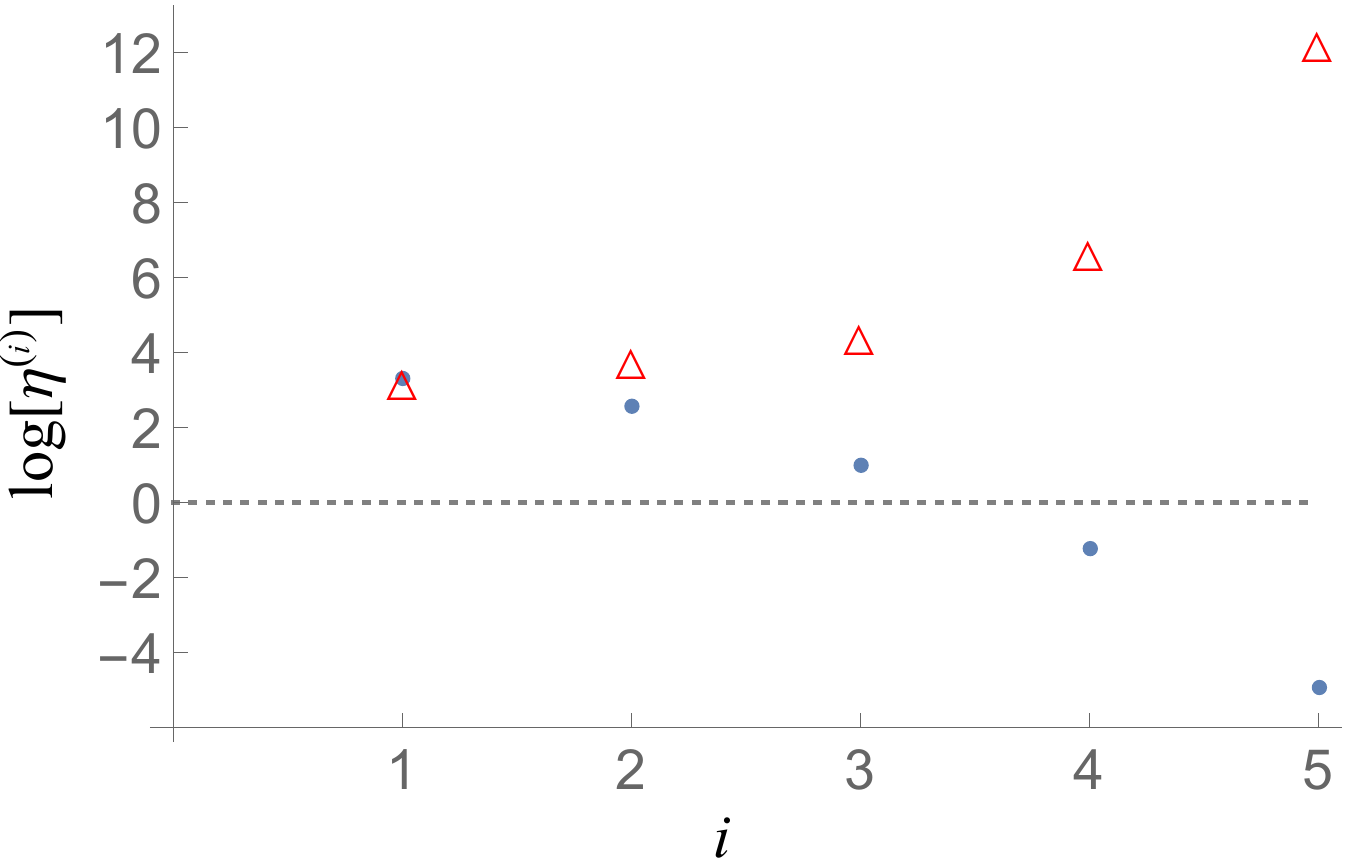}
\caption{(Color online) Logarithm (base 10) of the ratio of entanglement generation rate for the optimized buffer time protocol and the canonical protocol at different nesting levels for $p=0.02,\beta=0.2,p_T=1$ and two different values of $p_S=0.75$ (blue dots) and $p_S=0.5$ (red triangles). The buffer times of the nesting levels were approximately optimal in both cases.}
\label{fig:crossover}
\end{figure}

The manifold increase in the entanglement generation rate in the optimized buffer time protocol compared to canonical protocol is seen at all nesting levels if the swapping success probability $p_S$ is low, figure~\ref{fig:crossover}. On the other hand,  if $p_S$ is high then the canonical protocol can yield better rates for higher nesting levels. In OBP the probability factor on the right hand side of Eq.~(\ref{DEratei}) scales as, $p^{(i)}_{\textrm{out}}\sim (p_S)^i(p)^{2^i}$, for $p<<1$, whereas the time to obtain a state goes as, $n^{(i)}_{\textrm{in}}n^{(i)}_{\textrm{out}}2\tau^{(i)}_C= 2^{i+1}\tau_C$, for $n^{(i)}_{\textrm{out}}=2$ taken as an example. This implies a low probability of obtaining a remote entangled state per unit time. In the CP, the probability factor $p_S$ in the RHS of Eq.~(\ref{canDEratei}) is constant. However, the time to obtain a state, $(\prod_{j=1}^i\braket{k}_j)2\tau_C=\braket{k}_1(\braket{k}_2)^{i-1}2\tau_C$, can diverge much faster than that in the OBP, $2^{i+1}\tau_C$ in our example. This happens if the waiting time due to the swapping success probability, $\braket{k}_j\geq 2~ \forall j\geq 2$, which implies $p_S\leq 0.64$. In this case, the probability to obtain a remote entangled state per unit time in the CP can be even lower than that in the OBP. Moreover, the degradation in the fidelity, $\prod_{j=1}^{i}[\gamma^{X,(j)}]^{2^{i-j}}$, for $X=O,C$ in the two cases has maximum contribution from initial nesting levels. As discussed earlier the OBP yields a $~1/p$-factor increase of EGR at the first nesting level itself. Therefore the advantage due to OBP persists for the first few nesting levels till a crossover nesting level even if $p_S$ is high. At the crossover nesting level, the entanglement generation rate of the optimized buffer time protocol becomes equal to or less than that of the canonical protocol. If $p_S$ is low then the logarithm of the ratio of rates in the two protocols diverges with the nesting level. In practice, one can use the explicit expressions for the rates provided in Eq.~(\ref{DEratei}) and Eq.~(\ref{canDEratei}) to numerically evaluate the performance of the protocols at each nesting level. 

\begin{figure}
\centering
\includegraphics[height=6cm,width=.6\columnwidth]{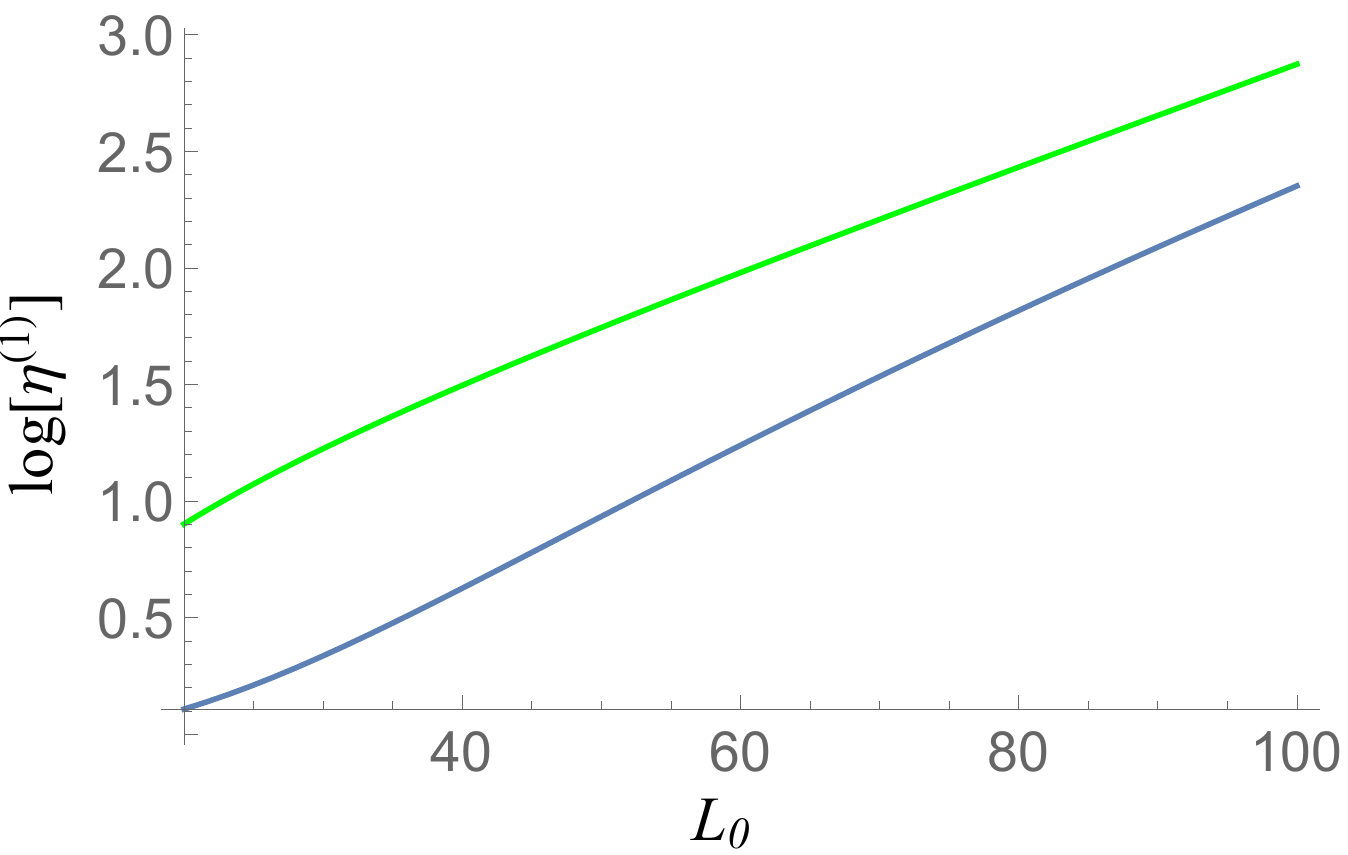}
\caption{(Color online) Logarithm (base 10) of the ratio of entanglement generation rate for the optimized buffer time protocol and the canonical protocol vs the distance between the repeater nodes, $L_0$, at the first nesting level. The green curve is obtained for memory lifetime $\tau_M=100~\mu$s and the blue curve for $\tau_M=1$ ms.}
\label{fig:ratio}
\end{figure}

The advantage of the optimized buffer time protocol can be observed by plotting the logarithm of the ratio of entanglement generation rates, $\log[\eta^{(i)}(p,\beta,n_{\textrm{opt}})]$, versus the distance between repeater nodes at the first nesting level as shown in figure (\ref{fig:ratio}). The repeater is placed midway, at a distance of $L_0$ from either end node, figure (\ref{fig:qlink}). We assume that the entanglement generation probability varies with the distance as $p=e^{-L_0/L_a}$ with the attenuation length $L_a=20$ kms. The memory quality parameter then varies as $\beta=e^{-(L_0/L_a)(L_a/c\tau_M)}$ where the speed of light in fiber is taken to be $c=2\times10^8$ ms$^{-1}$. We choose two different memory lifetimes with values of $\tau_M=100~\mu$s (green curve) and $\tau_M=1$ ms (blue curve). The latter shows that a hundred fold improvement in entanglement generation rates is obtained with 1 millisecond quantum memories when the repeater is at a distance of $100$ kms whereas the increase is even higher in the former case.

\section{Discussion and conclusion}
\label{sec:conc}

We presented a quantum repeater architecture and protocol that mitigates quantum memory decoherence. The protocol optimizes the buffer time of the quantum memories based on the operating point in parameter space. We showed the hierarchical optimization of the buffer time at all nesting levels. The resulting increase of entanglement generation rates by many orders of magnitude was demonstrated. Crucially, the improvement was achieved with state of art physical resources. For example, with current technology, entanglement generation probability of $p\simeq10^{-4}-10^{-5}$, memory lifetime of $\tau_M\simeq10^{-1}$ ms, swapping success probability $p_S\simeq 0.5$ and transfer probability $p_T\simeq1$ are feasible \cite{swap-experiment,quantummemories,seqquantumrepeater}. If repeater stations are spaced at intervals of $L_0=20$ km, corresponding to the attenuation length in optical fibers, the one-way classical communication time $\tau_C=L_0/(2\times 10^5~\textrm{km}~s^{-1})=10^{-1}$ ms equals the memory lifetime. The memory quality parameter then is $\beta=0.135$.  In this region of parameter space the optimized buffer time protocol yields $(10^4 -10^5)$ increase in the entanglement generation rate. The proposed optimized buffer time protocol performs particularly well in the technologically feasible parameter regions and could facilitate broad applications in the future development of quantum networks such as for interferometry \cite{santra_qtel} and secret sharing \cite{Hillery1999}.

%%%%%%%%%%%
\ack
This work was supported in part by the Office of the Secretary of Defense, Quantum Science and Engineering Program. L.J. acknowledges support from the ARL-CDQI (W911NF-15-2-0067, W911NF-18-2-0237), NSF (EFMA-1640959), and the Packard Foundation (2013-39273).

%%%%%%%%%%%%%%%%%%%%%%%
\appendix

\section{Derivation of the average state in the Optimized buffer-time protocol and the Canonical protocol}
\label{app:avstates}
The decoherence in a pair of identical quantum memories with a lifetime $\tau_M$, due to dephasing, occurs at the rate $2/\tau_M$. We will denote the phase decoherence superoperator by $D_t$. Thus for an initial stored state, $\rho_0=\ket{\psi^-}\bra{\psi^-}$, we have
\begin{equation}
\rho(t)=D_t(\rho_0)=P^-(t)\rho^+ + P^+(t)\rho^-,
\end{equation}
with $P^{\pm}:=(1\pm e^{-2t/\tau_M})/2$.

We assume that our heralded scheme of entanglement swapping succeeds with a probability $p_S$. Under the swapping operation $\hat{S}$ for a pair of 2-qubit states $\rho(t_1)$ and $\rho(t_2)$ stored in the two pairs of memories for times $t_1$ and $t_2$ we have the output state conditioned on heralding to be
\begin{eqnarray}
\hat{S}[\rho(t_1),\rho(t_2)]=\hat{S}[(P^-(t_1)\rho^+ + P^+(t_1)\rho^-)(P^-(t_2)\rho^+ + P^+(t_2)\rho^-)]\nonumber\\
=P^-(t_1)P^-(t_2)\hat{S}[\rho^+,\rho^+]+P^-(t_1)P^+(t_2)\hat{S}[\rho^+,\rho^-]\nonumber\\
~~+P^+(t_1)P^-(t_2)\hat{S}[\rho^-,\rho^+]+P^+(t_1)P^+(t_2)\hat{S}[\rho^-,\rho^-]\nonumber\\
=(P^-(t_1)P^-(t_2)+P^+(t_1)P^+(t_2))\rho^-+(P^+(t_1)P^-(t_2)+P^-(t_1)P^+(t_2))\rho^+\nonumber\\
=P^+(t_1+t_2)\rho^-+P^-(t_1+t_2)\rho^+,
\label{swapstate}
\end{eqnarray}
where we have used the linearity of the swap operation in the second line and the equalities $\hat{S}[\rho^+,\rho^+]=\hat{S}[\rho^-,\rho^-]=\rho^-$, $\hat{S}[\rho^+,\rho^-]=\hat{S}[\rho^-,\rho^+]=\rho^+$.

The probabilistic process of charging a pair of memories with the state $\ket{\psi^-}$ succeeds at some step number $k$. This can happen for possibly different step numbers $k_1,k_2$ for the two pairs of memories shown in figure (1) of the main text. Assuming $k_1\leq k_2$, the latter pair of memories still stores the state for a time $t_2=\tau_C$, i.e., for one classical communication time. The earlier charged pair stores the state for a time $t_1=(k_2-k_1)2\tau_C+\tau_C$. Therefore, the correspondence of storage times to step numbers is, 
\begin{eqnarray}
t_1\to(k_2-k_1)2\tau_C+\tau_C\nonumber\\
t_2\to\tau_C.
\end{eqnarray}
Thus, $(t_1+t_2)\to (|k_2-k_1|+1)2\tau_C$ for all $k_1,k_2$. 

In the OBP the average accessed state $\rho^O$ is a probabilistically weighted sum of states obtained after swapping, the states that have been stored in the two pairs of quantum memories, subject to the condition on the charging step numbers $|k_2-k_1|\leq n$. The probability distribution of successful entanglement generation has the form $\mathbb{P}(k_1,k_2)=(1-p)^{k_1+k_2-2}p^2$. The form of the remote entangled state $\rho^S(k_1,k_2)$ after the swap is given by Eq.~(\ref{swapstate}) with $(t_1+t_2)\to (|k_2-k_1|+1)2\tau_C$ for all $k_1,k_2$. One also needs to account for the decoherence suffered in the two remote memories after a swapped state is obtained untill the memories are refreshed after $n$ cycle times. This state obtained after the swap, Eq.~(\ref{swapstate}), further decoheres in the remote quantum memories at the two ends for a time, $t_n=\{2(n-\textrm{max}(k_1,k_2))+1\}\tau_C$, and leads to the state $\rho^S_n(k_1,k_2)=D_{t_n}(\rho^S(k_1,k_2))$. Thus the average state in the optimized  protocol is given by
\begin{eqnarray}
\rho^O=\frac{\sum_{k_1=1,k_2=1}^{n,n} \mathbb{P}(k_1,k_2)\rho^S_n(k_1,k_2)}{(1-(1-p)^n)^2}\nonumber\\
=\frac{1+\gamma^O(p,\beta,n)}{2}\rho^-+\frac{1-\gamma^O(p,\beta,n)}{2}\rho^+,
\label{optstate}
\end{eqnarray}
where $\gamma^O(p,\beta,n)$ is given by
\begin{equation}
\gamma^O(p,\beta,n) = \beta^3\frac{p}{(1-(1-p)^n)^2}\frac{f(p,\beta,n)}{(\beta^2-q)(\beta^2-q^2)},
\label{eq:gammaopt}
\end{equation}
with $f(p,\beta,n) = q^{2n}(\beta^2+q(1-q-\beta^2))+\beta^{2n}(2q^{n+2}-q^2+\beta^2-2q^n\beta^2+q(\beta^2-1))$, $q=(1-p)$, and  $\beta=e^{-2\tau_C/\tau_M}$. Note that the function $f(p,\beta,n)$ has $q=\beta,\beta^2$ as roots and thus the factors in the denominator of Eq.~(\ref{eq:gammaopt}) do not cause any singular behavior. For perfect quantum memories $\beta=1$ so that $\gamma^O(p,\beta,n)=0$. For $p=1$, $\gamma^O(p,\beta,n)=\beta^{2n+1}$ which is maximum for $n=1$ and $\beta<1$.

The state in Eq.~(\ref{optstate}) is the output of the first nesting level and is denoted as $\rho^{O,(1)}$. Second nesting level receives the average output state from the first nesting level and outputs a probabilistically weighted average state with the probability defined by the input probability for the second level. The probability distribution at any nesting level, $i$, is given by
\begin{equation}
\mathbb{P}^{O,(i)}(k_1,k_2)=(p^{(i)}_{\textrm{in}})^2(q^{(i)}_{\textrm{in}})^{k_1+k_2-2} 
\label{probgreedyj}
\end{equation}
with $q^{(i)}_{\textrm{in}}=(1-p^{(i)}_{\textrm{in}}$). The average state obtained in the optimistic protocol at any nesting level depends on the sequence of values of the parameter sets $\bar{\tau}_C=\{\tau^{(1)}_C,\tau^{(2)}_C,...,\tau^{(i)}_C\}$,
$\bar{\beta}=\{\beta^{(1)},\beta^{(2)},...,\beta^{(i)}\}$, $\bar{n}_{\textrm{in}}=\{n^{(1)}_{\textrm{in}},n^{(2)}_{\textrm{in}},...,n^{(i)}_{\textrm{in}}\}$, the initial charging success probability $p$, the swapping success probability $p_S$, and the transfer success probability $p_T$ . This state is obtained by iterating the process of averaging over the states received from the previous nesting level and normalizing by the appropriate probability normalization factor resulting in
\begin{equation}
\rho^{O,(i)}_{\textrm{out}}=\frac{1+\prod_{j=1}^i[\gamma^{O,(j)}]^{2^{i-j}}}{2}\rho^- +\frac{1-\prod_{j=1}^i[\gamma^{O,(j)}]^{2^{i-j}}}{2}\rho^+.
\label{eq:rhonestleveli}
\end{equation}
In the above expression the explicit dependence of $\gamma^{O,(j)}$ on the relevant parameter values are suppressed for brevity. The actual expression for which is
\begin{equation}
\gamma^{O,(j)}=\mathcal{N}^{-1}\sum_{k_1,k_2=1}^{n^{(j)}_{\textrm{out}}}\mathbb{P}^{O,(j)}(k_1,k_2)[\beta^{(j)}]^{n^{(j)}_{\textrm{in}}\{2(n^{(j)}_{\textrm{out}}-k_1)+2\}+1},
\end{equation}
where $\mathcal{N}=\sum_{k_1,k_2=1}^{n^{(j)}_{\textrm{out}}}\mathbb{P}^{(j)}(k_1,k_2)$. 

In the CP, the average obtained state is a probabilistically weighted sum of states over all possible storage times in the two pairs of memories. The state obtained after the swap operation is given by Eq.~(\ref{swapstate}) with $(t_1+t_2)\to (|k_2-k_1|+1)2\tau_C$ for all $k_1,k_2$. The obtained swapped state decoheres during time $\tau_C$ in the two remote memories resulting in the state, $\rho^S_r(k_1,k_2)=D_{\tau_C}(\rho^S(k_1,k_2))$, since the results of the swap operation have to be communicated to the end nodes. Therefore, the CP average state is 
\begin{eqnarray}
\rho^C=\sum_{k_1=1,k_2=1}^{\infty,\infty} \mathbb{P}(k_1,k_2)\rho^S_r(k_1,k_2)\nonumber\\
=\frac{1+\gamma^C(p,\beta)}{2}\rho^-+\frac{1-\gamma^C(p,\beta)}{2}\rho^+ \, ,
\label{expectedstategreedy}
\end{eqnarray}
where the function $\gamma^C(p,\beta)$ is given by
\begin{equation}
\gamma^C(p,\beta) = \beta^3p^2\frac{(1-\beta^2 q)+2\beta^2 q}{(1-q^2)(1-\beta^2 q)}
\label{gammagreedy}
\end{equation}
with $\beta=e^{-2\tau_C/\tau_M}$ and $q=1-p$. Again, for $\beta=1$, $\gamma^C(p,\beta)=1$ and for $p=1$, $\gamma^C(p,\beta)=\beta^3$. The expected number of steps to obtain an entangled state in both involved segments is given by
\begin{eqnarray}
\braket{k}=\sum_{k=1}^\infty kp^2q^{2(k-1)}+2\sum_{k_1=1}^\infty p^2q^{k_1-2}(\sum_{k_2=k_1+1}^{\infty}k_2q^{k_2})\nonumber\\
=\frac{(3-2p)}{p(2-p)} \,.
\end{eqnarray}
The average state after any nesting level in the CP is obtained by iterating the averaging procedure at every nesting level. The probability distribution at the first nesting level is determined by the entanglement generation probability. For the second and higher nesting levels the probability distribution is determined by the swapping success probability. Thus,
\begin{eqnarray}
\mathbb{P}^{C,(1)}(k_1,k_2)=p^2q^{k_1+k_2-2} \,, \nonumber\\
\mathbb{P}^{C,(i)}(k_1,k_2)=p_S^2q_S^{k_1+k_2-2} \,.
\end{eqnarray}
By iterating the averaging procedure at each nesting level using the above probability distributions we get the form of the average state for any nesting level
\begin{equation}
\rho^{C,(i)}_{\textrm{out}}=\frac{1+\prod_{j=1}^i[\gamma^{C,(j)}]^{2^{i-j}}}{2}\rho^- +\frac{1-\prod_{j=1}^i[\gamma^{C,(j)}]^{2^{i-j}}}{2}\rho^+,
\label{eq:canrhonestleveli}
\end{equation}
where 
\begin{equation}
\gamma^{C,(j)}=\sum_{k_1,k_2=1}^{\infty,\infty}\mathbb{P}^{C,(j)}(k_1,k_2)[\beta^{(j)}]^{3}[\beta^{(1)}]^{2\prod_{l=0}^{(j-1)}\braket{k}_l(k_2-k_1)},
\end{equation}
with $\braket{k}_0=1$, $\braket{k}_1=(3-2p)/p(2-p)$, $\braket{k}_l=(3-2p_S)/p_S(2-p_S)$ for $l\geq 2$.

In both the OBP and CP, if the initial entangled state across the elementary segments has a fidelity of $f$, i.e., $\rho_0=f\rho^-+(1-f)\rho^+$, then the output states Eq.~(\ref{eq:rhonestleveli}) and Eq.~(\ref{eq:canrhonestleveli}) include the fidelity factor $f$ in their coefficients. Thus for the OBP  $\gamma^{O,(1)}\to f\gamma^{O,(1)}$, while for the CP $\gamma^{C,(1)}\to f\gamma^{C,(1)}$. Our results reman unchanged for any value of the initial fidelity.

 For a repeater with $N_m$ nesting levels, OBP requires $2(2^{N_m+2}-2)$ memories whereas CP requires $2^{N_m+2}$ quantum memories. Thus, the OBP requires at most twice as many quantum memories as the CP. The EGR per memory used is still higher by orders of magnitude in the OBP in the relevant regions of parameter space.

\section{Ratio of rates and optimal wait-window size}
\label{app:nopt}

\begin{figure}
\centering
\includegraphics[height=6cm,width=.6\columnwidth]{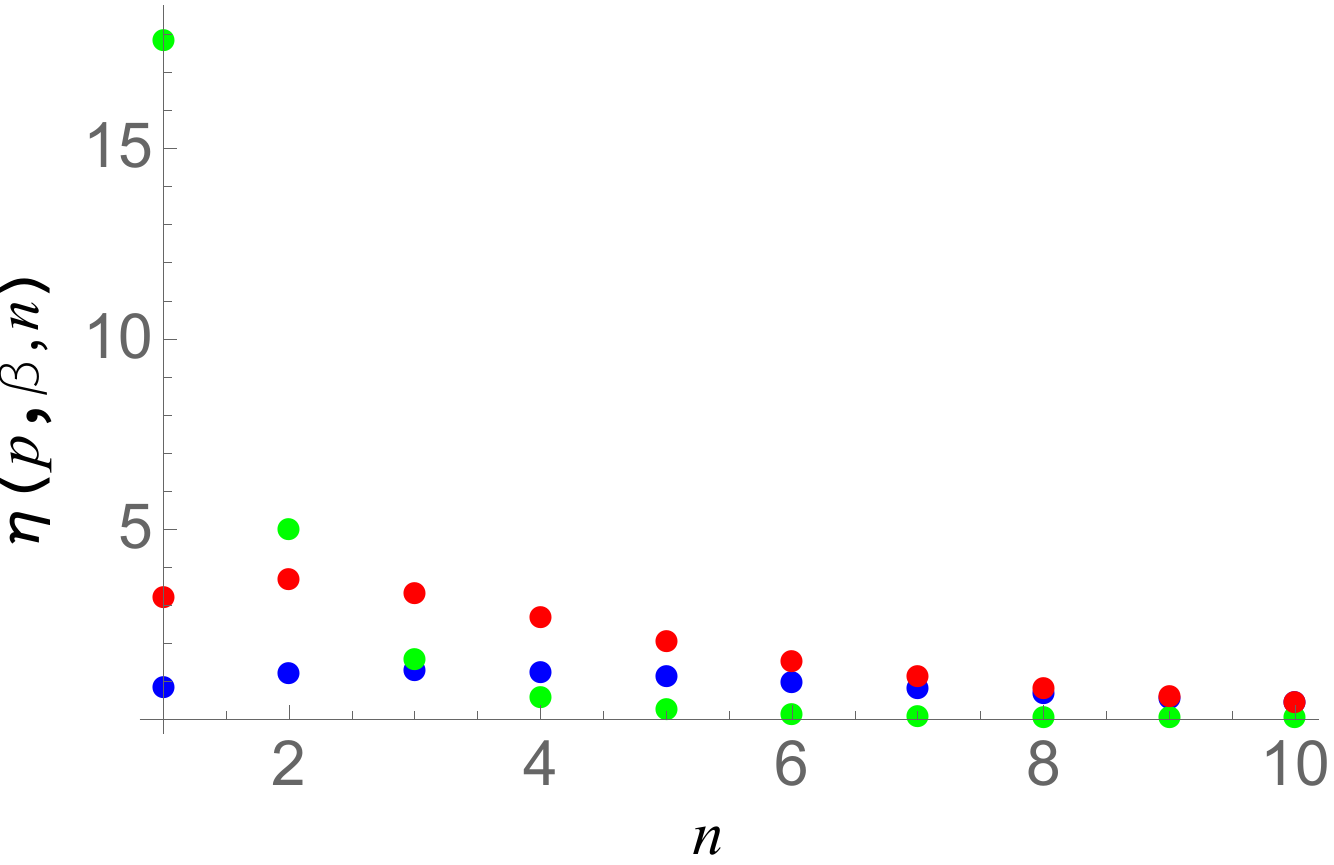}
\caption{The optimal wait-window size in the optimized buffer time protocol depends on the location in parameter space. Shown here is the ratio of entanglement generation rates for $(p,\beta)$ = (0.1,0.9) in blue, (0.1,0.4) in green and (0.05,0.8) in red. The sequences of blue, red and green dots have its maximum at $n_{\textrm{opt}}=3,2,1$ respectively.}
\label{fig:optimaln1}
\end{figure}

The OBP provides higher entanglement generation rates than the CP for most regions of the $(p,\beta)$-parameter space, as shown in figure~2 of the main text. The size of the optimal wait-window in terms of cycle time depends on the operating point in the $(p,\beta)$ parameter space, figure~(\ref{fig:optimaln1}).

\begin{figure}[h]
\centering
\includegraphics[height=6cm,width=.6\columnwidth]{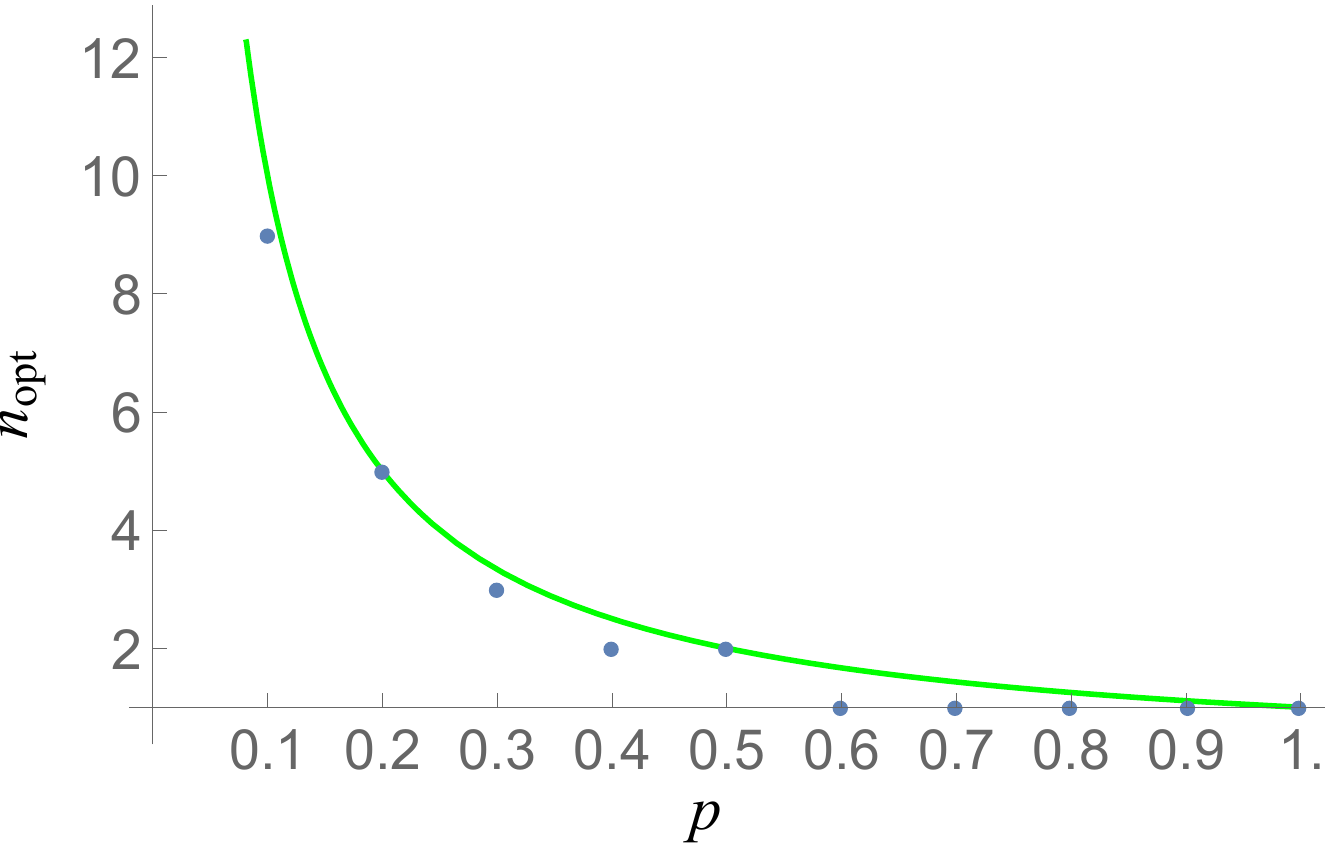}
\caption{Scaling of the optimal wait-window size $n_{\textrm{opt}}$ (blue dots) with the charging success probability $p$ for a fixed value of $\beta=0.99$. A scaling of $n_{\textrm{opt}}\sim 1/p$ is observed as $p\to 0$ while $n_{\textrm{opt}}=1$ for $p>.5$.}
\label{fig:optimistic-optimaln1}
\end{figure}

\begin{figure}[h]
\centering
\includegraphics[height=6cm,width=.6\columnwidth]{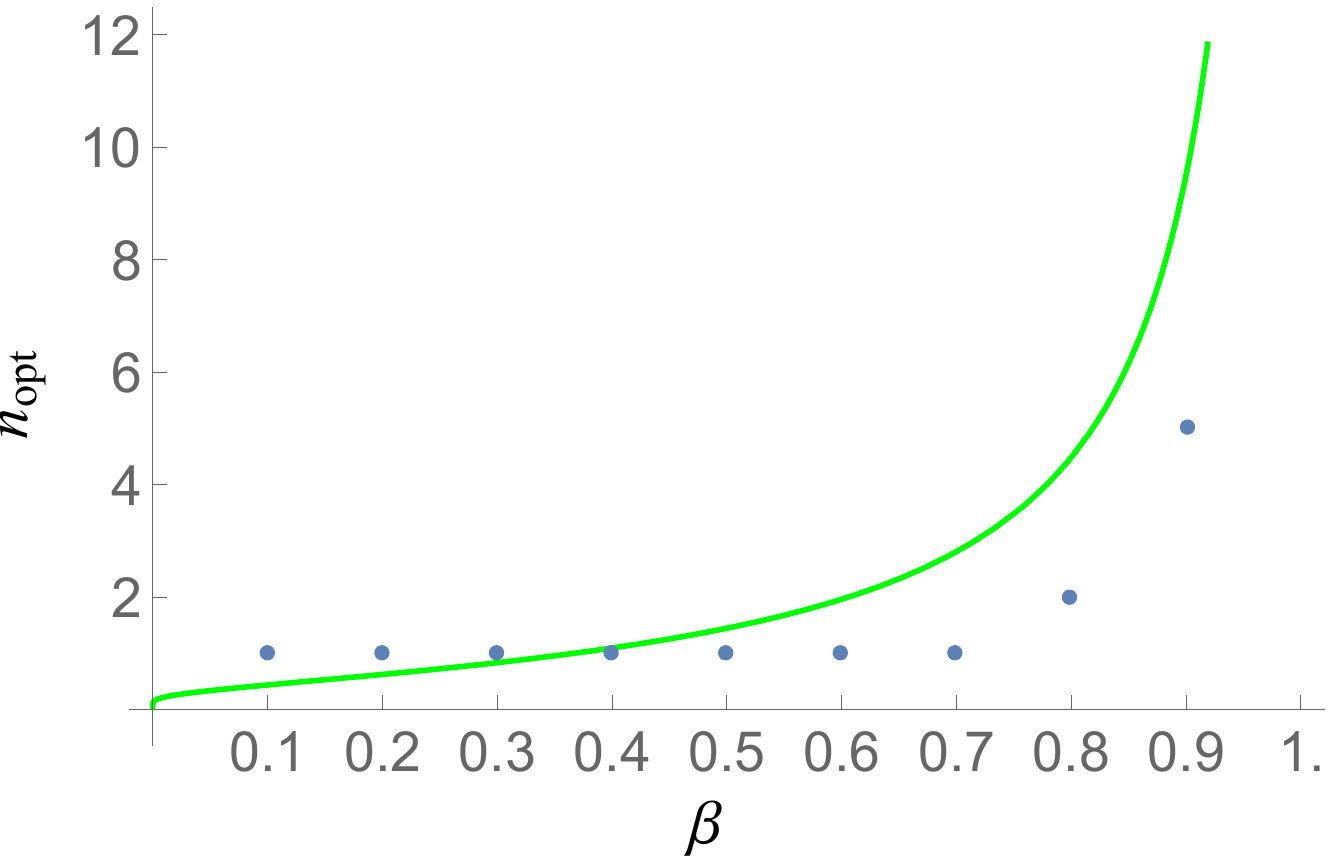}
\caption{Scaling of the optimal wait-window size $n_{\textrm{opt}}$ (blue dots) with the memory parameter $\beta$ for a fixed value of $p=0.01$. A scaling of $n_{\textrm{opt}}\sim -1/\log(\beta)$ is observed as $\beta\to 1$ while $n_{\textrm{opt}}=1$ for $\beta\leq.7$.}
\label{fig:optimistic-optimaln2}
\end{figure}

We identify several regions:
\protect{\begin{itemize}
\item Long-lived quantum memories, low entanglement generation probabilities ($p\to 0,~\beta\to 1$). In this region, figure~(\ref{fig:optimistic-optimaln1}) suggests the scaling of $n_{\textrm{opt}}\sim1/p$.
\item Short-lived quantum memories, high entanglement generation probabilities ($p\to 1,~\beta\to 0$). In this region, the best schedule of course is the rapid reset strategy with $n_{\textrm{opt}}=1$.
\item Short-lived memories, low entanglement generation probabilities ($p\to 0,~\beta\to 0$). In this region, figure~(\ref{fig:optimistic-optimaln2}) suggests that rapid resetting with $n_{\textrm{opt}}=1$ constitutes the best schedule. 
\item Long-lived quantum memories, high entanglement generation probabilities ($p\to 1,~\beta\to 1$). In this region, the best schedule also turns out to be the rapid reset strategy $n_{\textrm{opt}}=1$. For $\beta\simeq 1$ the CP provides better entanglement generation rates than the OBP. 
\end{itemize}}

\section{Ratio of entanglement generation rates for nesting level N=1}
\label{app:rateratio}
\begin{figure}[h]
\centering
\includegraphics[height=6cm,width=.6\columnwidth]{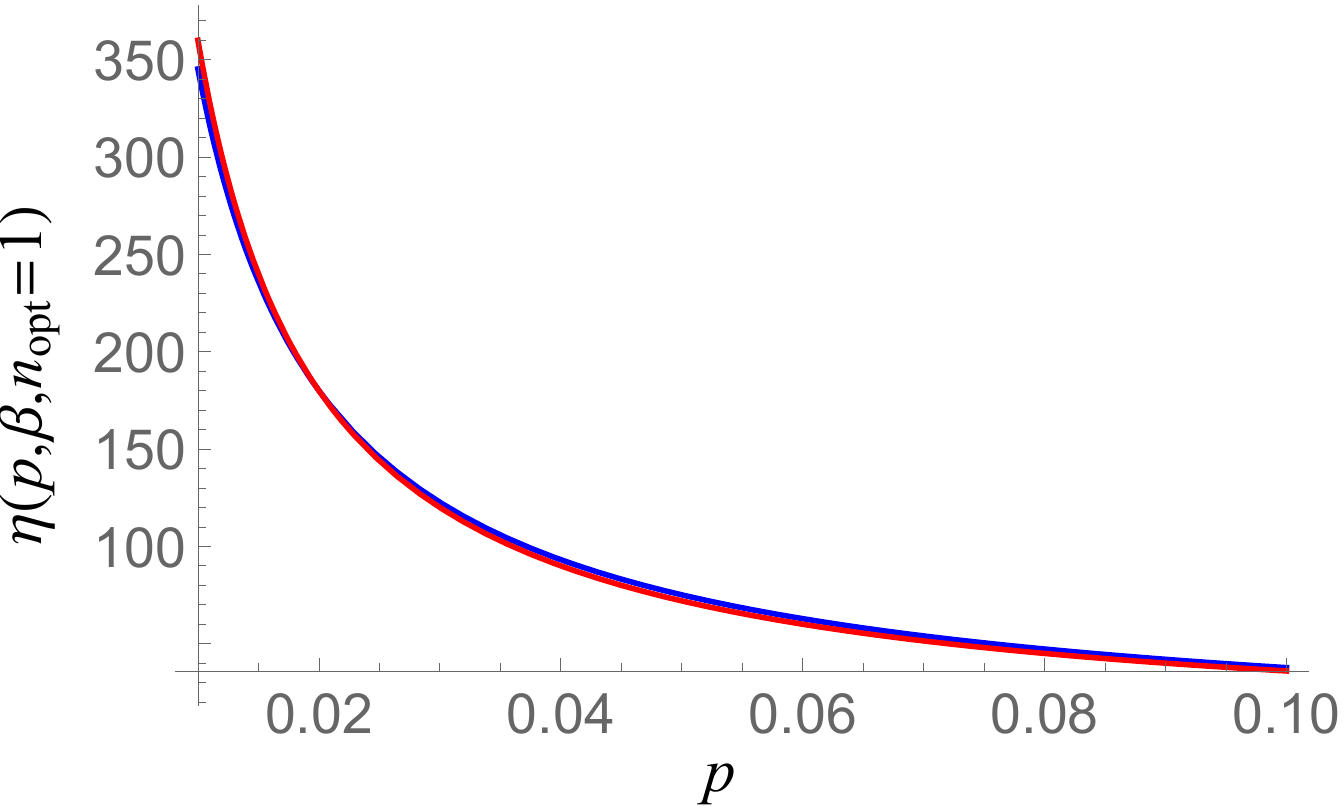}
\caption{Scaling of the entanglement generation rate ratio using the optimized buffer time protocol and the canonical protocol: blue curve - the exact rate expressions from (\ref{ratecomp1}) for $\beta=.1$ and $p\in[1\%,10\%]$ ; red curve  - approximation as $3.6/p$.}
\label{fig:ratioscaling1}
\end{figure}

The OBP and the CP can be compared with respect to the entanglement generation rate which is the maximum rate of distillable entanglement. Here we compare the distillable entanglement generation rates in the two protocols at the first nesting level, $i=1$, in the $p\to0,\beta\to 0$ region. We find approximations for the rates
\begin{eqnarray}
R^C_{DE}(p,\beta)=\frac{p_S}{\tau_C}\frac{p(2-p)}{3-2p}H[x_1(p,\beta)],\nonumber\\
R^O_{DE}(p,\beta,n)=\frac{p_S}{\tau_C}\frac{(1-(1-p)^n)^2}{n}H[x_2(p,\beta,n)],
\label{ratecomp1}
\end{eqnarray}
where $x_2=(1+\sqrt{1-(\gamma^O(p,\beta,n))^2})/2$, and $x_1=(1+\sqrt{1-(\gamma^C(p,\beta))^2})/2$. In the $p\to0,\beta\to 0$ region both $\gamma^C,\gamma^O\ll 1$ so that $x_1(p,\beta)\approx 1-(\gamma^C(p,\beta))^2/4$, $x_2(p,\beta,n)\approx 1-(\gamma^O(p,\beta,n))^2/4$, $\gamma^C(p,\beta)\approx \beta^3p/2$, $\gamma^O(p,\beta,n)\approx \beta^3/n^2$. We now use the property of binary entropy that, $H(1-x)=H(x), 0\leq x\leq 1$ and a small $x$ approximation, $H(x)\approx x\log_2(e/x), x\to 0$. Further, we approximate $p(2-p)/3-2p\approx 2p/3$ and $(1-(1-p)^n)^2/n\approx np^2$. We also know from our numerical investigations (and analytical results) that in this region $n_{\textrm{opt}}=1$. Putting all this together we get
\begin{eqnarray}
\eta (p,\beta,n_{\textrm{opt}}=1) =
\frac{R^O_{DE}(p,\beta,n_{\textrm{opt}}=1)}{R^C_{DE}(p,\beta)}\nonumber\\
\approx\frac{6}{p}\frac{\log_2(4e)-6\log_2(\beta)}{\log_2(16e)-6\log_2(\beta)-2\log_2(p)}.
\end{eqnarray}
For $p,\beta$ values of technological relevance the above ratio is well approximated as $1/p$, shown in figure~(\ref{fig:ratioscaling1}).

\section*{References}
\bibliographystyle{iopart-num}
\bibliography{refs-greedy1}
\end{document}